\documentclass[preprint,12pt, a4paper]{elsarticle}

\usepackage{amssymb}
\usepackage{hyperref}
\usepackage{tabularx}
\usepackage{adjustbox}
 \usepackage{multirow}
 \usepackage{xcolor}
 \newcommand{\code}[1]{\texttt{#1}}
\journal{SoftwareX}

\begin{document}
\renewcommand{\labelenumii}{\arabic{enumi}.\arabic{enumii}}

\begin{frontmatter}
\title{A Kokkos-Accelerated Moment Tensor Potential Implementation for LAMMPS}


\author[label1,label5]{Zijian Meng}
\author[label1]{Karim Zongo}
\author[label4]{Edmanuel Torres}
\author[label4]{Christopher Maxwell}
\author[label2]{Ryan Eric Grant}
\author[label1]{Laurent Karim Béland}
\address[label1]{Department of Mechanical and Materials Engineering, Queen's University Kingston, Ontario, Canada, K7L 3N6}
\address[label2]{Department of Electrical and Computer Engineering, Queen's University Kingston, Ontario, Canada, K7L 3N6}
\address[label4]{Canadian Nuclear Laboratories  Chalk River, Ontario, Canada}
\address[label5]{17zjm1@queensu.ca; contact@richardzjm.com}

\begin{abstract}
We present a Kokkos-accelerated implementation of the Moment Tensor Potential (MTP) for LAMMPS, designed to improve both computational performance and portability across CPUs and GPUs. This package introduces an optimized CPU variant—achieving up to 2× speedups over existing implementations—and two new GPU variants: a thread-parallel version for large-scale simulations and a block-parallel version optimized for smaller systems. It supports three core functionalities: standard inference, configuration-mode active learning, and neighborhood-mode active learning. Benchmarks and case studies demonstrate efficient scaling to million-atom systems, substantially extending accessible length and time scales while preserving the MTP’s near-quantum accuracy and native support for uncertainty quantification.
\end{abstract}


\end{frontmatter}

\section{Motivation and significance}

Atomistic simulations have become a vital complement to experimental methods in materials discovery and characterization \cite{lesar2013introduction,gowthaman2023review,hollingsworth2018molecular}. These simulations rely on interatomic potentials—models of atomic interactions—to compute energies and forces. Traditional potentials are computationally efficient but often lack the flexibility needed for high-fidelity predictions, while quantum mechanical methods such as density functional theory (DFT) offer greater accuracy at substantially higher computational cost. Machine learning interatomic potentials (MLIPs) have emerged as a compelling alternative, providing a systematic framework to improve accuracy by increasing model complexity, thereby enabling better control over the cost-accuracy trade-off than conventional approaches~\cite{engkvist2000accurate, deringer2019machine}. However, the computational demands of high-capacity MLIPs remain a challenge, especially for large-scale or long-timescale simulations.

The rise of MLIPs has paralleled growing demand for computational power, making hardware accelerators—particularly GPUs—essential for large-scale simulations. At the same time, the landscape of high-performance computing (HPC) has become increasingly heterogeneous. Leading supercomputers such as LUMI, Frontier, and the newly commissioned El Capitan employ AMD Instinct GPUs (MI250X, MI300A)~\cite{markomanolis2022evaluating}, while Aurora relies on Intel Max GPUs. However, fully leveraging these diverse architectures often requires adopting distinct programming models, vendor-specific frameworks, or low-level languages, posing a significant challenge for portability and maintainability.

LAMMPS—Large-scale Atomic/Molecular Massively Parallel Simulator—a widely used classical atomistic simulation engine~\cite{plimpton1995fast,thompson2022lammps}, supports Kokkos-based acceleration to enable performance portability across diverse hardware platforms. Kokkos provides a unified abstraction for parallel execution and data management~\cite{9485033,CarterEdwards20143202}, allowing a single implementation to target CPUs, NVIDIA GPUs, AMD GPUs, and other architectures without sacrificing performance. Several interatomic potentials—such as the Tabulated Gaussian Approximation Potential (tabGAP) \cite{byggmastar2022simple,luo2024interatomic}, the Spectral Neighbor Analysis Potential (SNAP) \cite{thompson2015spectral}, and the Atomic Cluster Expansion (ACE/PACE) \cite{drautz2019atomic,lysogorskiy2021performant}—have been implemented in this framework, and have demonstrated scalability to billion-atom systems and nanosecond timescales  \cite{nguyen2021billion}. These capabilities bring large-scale, high-fidelity simulations closer to experimentally relevant conditions in both space and time.

We extend the LAMMPS Kokkos package to support the Moment Tensor Potential (MTP)\cite{shapeev2016moment}, including both inference and uncertainty quantification for active learning. MTP is one of the most widely used machine learning interatomic potentials, with demonstrated success across metals, semiconductors, and multicomponent systems \cite{podryabinkin2017active,novikov2020mlip,zongo2024unified,luo2023set,sun2024interatomic,meziere2023accelerating,kwon2023accurate,wang2025efficient,novikov2022magnetic,attarian2022thermophysical,novoselov2019moment, qi2023machine,chen2023development,poul2023systematic}. It offers a strong balance of computational efficiency, systematic improvability, and built-in support for active learning via extrapolation grades based on D-optimality~\cite{kiefer1959optimum} and the MaxVol algorithm~\cite{goreinov2010find}. While newer formalisms such as ACE have outperformed MTP on recent Pareto fronts of accuracy versus cost~\cite{leimeroth2025machine}, ongoing work has shown that careful adjustments to the basis construction can further improve its performance and expressiveness \cite{wang2025efficient}. Integrating MTP into the Kokkos framework within LAMMPS enables scalable, portable deployment across GPU architectures, significantly broadening the range of accessible simulation sizes and timescales.

The software is available on GitHub\footnote{\url{https://github.com/RichardZJM/lammps-mtp-kokkos}} \cite{meng_2025_lammps}.

\section{Software description}
\subsection{Software architecture}
We introduce nine new implementations of the Moment Tensor Potential, each optimized for different simulation conditions to provide flexibility across a wide range of system sizes and hardware configurations. These cover three core use cases: inference, active learning in configuration mode, and active learning in neighborhood mode. For each use case, we provide three implementations: (1) a further optimized CPU version (non-Kokkos) that improves upon the original MLIP-3 package~\cite{podryabinkin2023mlip}; (2) a thread-parallel GPU variant designed for large-scale simulations (typically $\gtrsim$50,000 atoms per GPU); and (3) a block-parallel GPU variant optimized for smaller simulations (typically $\gtrsim$2000 atoms per GPU). The block-parallel version, denoted with \code{small} in the LAMMPS \code{pair\_style}, exposes additional fine-grained parallelism, which can improve performance at small-to-intermediate system sizes but may reduce peak throughput. Each variant follows LAMMPS \code{pair\_style} naming conventions, as summarized in Table~\ref{tab:modes}.

\begin{table}[htbp]
    \centering
    \caption{Each MTP variation and its LAMMPS \code{pair\_style} identifier. }
    \label{tab:modes}
    \begin{tabular}{c|l|c|c}
        \hline
        \multicolumn{2}{c|}{\multirow{2}{*}{Platform}} & \multicolumn{2}{c}{Use Cases} \\
        \cline{3-4}
        \multicolumn{2}{c|}{} & \multicolumn{1}{c|}{Inference} & Active Learning, Both Modes \\
        \hline
        \multicolumn{2}{c|}{CPU} & \code{mtp} & \code{mtp/extrapolation} \\
        \hline
        \multirow{2}{*}{GPU} & Thread-Parallel & \code{mtp/kk} & \code{mtp/extrapolation/kk} \\
        & Block-Parallel & \code{mtp/small/kk} & \code{mtp/extrapolation/small/kk} \\
        \hline
    \end{tabular}
\end{table}

Users should first select the desired use case. Inference computes energies, forces, and stresses during standard molecular dynamics simulations. Active learning performs the same calculations while also evaluating the extrapolation grade at user-defined intervals, enabling on-the-fly model improvement. Configuration mode and neighborhood mode refer to two distinct strategies for computing this grade. As in MLIP-3, both modes support a selection threshold—above which configurations are written to disk—and a break threshold, above which the simulation is halted. Once the use case is chosen, the appropriate implementation (CPU, thread-parallel GPU, or block-parallel GPU) can be selected. For GPU usage, we recommend short single-GPU trial runs of both variants, as relative performance depends strongly on the MTP parameters and the underlying hardware.

\subsection{Software functionalities}
\begin{figure}[]
	\centering 
	\includegraphics[width=\linewidth]{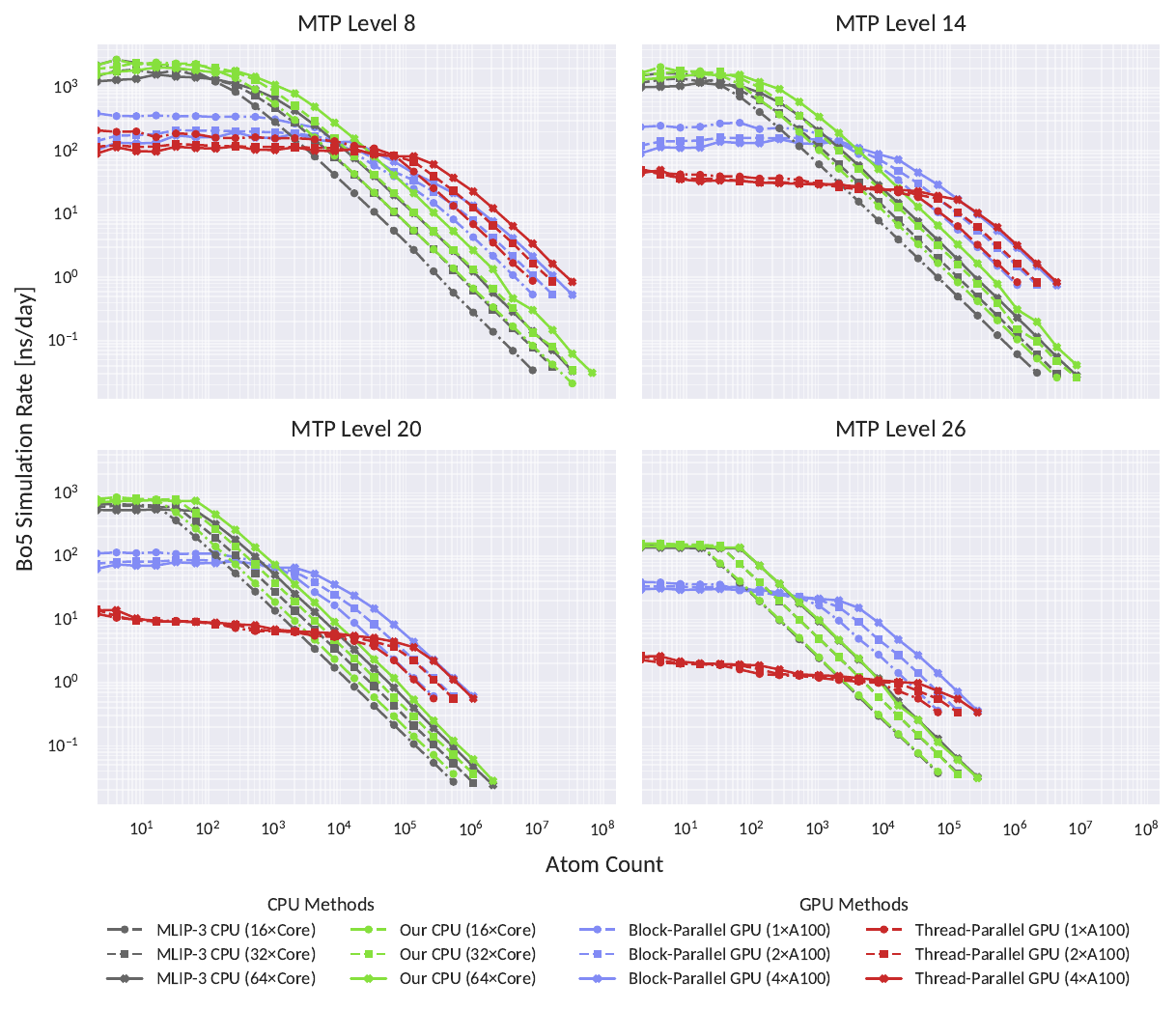}	
	\caption{A log-log plot of the inference simulation rate versus the atom count of several MTP implementations on various hardware for selected MTP levels. Separate simulations are performed for 100 timesteps for each atom count and method, and the best of five (Bo5) simulation rate is reported. 1 fs timestep is used.} 
	\label{fig:perf_infer}%
     \hspace{3cm}
    \includegraphics[width=0.83\linewidth]{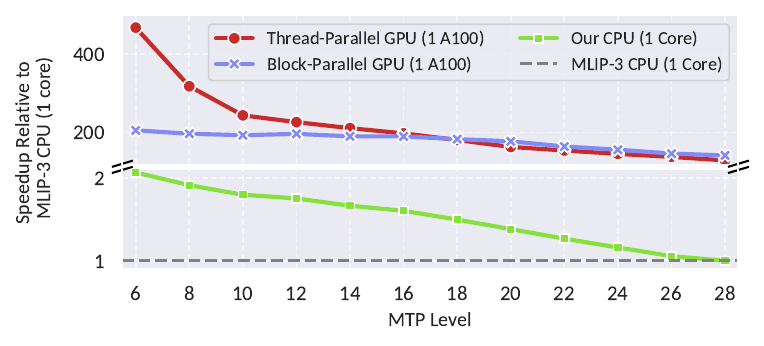}
	\caption{The inference speedups (relative maximum throughput) over the previous MLIP-3 implementation.} 
	\label{fig:speedup}%
\end{figure}

The main contributions of the software are improved CPU performance and new GPU capabilities, while fully preserving the core functionality of the MTP as described in previous works \cite{shapeev2016moment, podryabinkin2023mlip}. To evaluate these improvements, we benchmarked both weak and strong scaling on the Digital Research Alliance of Canada's Narval HPC cluster. Narval's CPU nodes are equipped with 2× AMD EPYC™ 7532 processors (32 cores each), and its GPU nodes with 4× NVIDIA A100 SXM4 (40 GB). MTP models are characterized by their “level”, which exponentially scales the number of basis functions—and thus the computational cost. Figure \ref{fig:perf_infer} presents weak and strong scaling results across a range of MTP levels, using a quarter, half, and full Narval node (CPU and GPU). Except for the MTP level, all benchmarks use default hyperparameters (e.g., cutoff radius) for a bulk simulation of unstrained solid potassium. The complete LAMMPS input script and full benchmarking data are provided in a separate repository \cite{zijian_2025_kokkos}.

\begin{figure}[]
	\centering 
	\includegraphics[width=\linewidth]{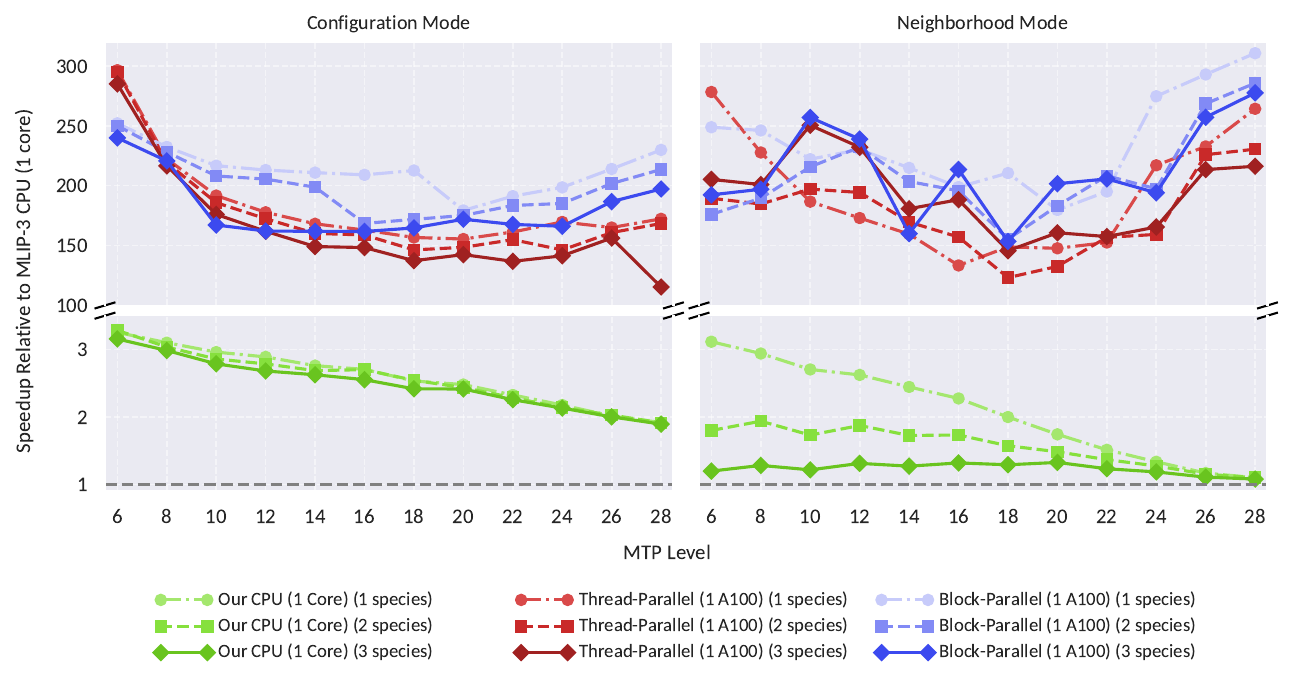}
	\caption{The active learning speedups (relative maximum throughput) for both configuration and neighborhood mode over previous the MLIP-3 implementation.} 
	\label{fig:al_speedup}%
\end{figure}

For each MTP level, we evaluated the maximum throughput—measured in atom-timesteps per wall-time second—across all tested atom counts and benchmark trials, and report relative speedups. Inference speedups compared to the original MLIP-3 implementation are shown in Figure~\ref{fig:speedup}, using a single CPU core and a single NVIDIA A100 GPU. Figure~\ref{fig:al_speedup} presents corresponding speedups for active learning in both configuration and neighborhood modes, evaluated across systems containing one, two, or three atomic species.

Notably, the crossover point at which the thread-parallel variant outperforms the block-parallel variant depends strongly on both the atom count and MTP level, as well as the underlying hardware. Overall, the observed GPU speedups are comparable to the acceleration achieved by the existing Kokkos implementation of the ACE potential when comparing a single A100 GPU to a single CPU core~\cite{leimeroth2025machine}.


 \subsection{Sample Code Snippets}
As with other interatomic potentials in LAMMPS, MTP variants are specified using the \code{pair\_style} command, followed by the path to the MTP potential file. These files are backward-compatible with the MLIP-3 format. For GPU variants, a chunk size must be provided using the \code{chunksize} keyword to manage memory usage. If the total number of atoms exceeds the specified chunk size, the simulation proceeds in multiple chunks. For optimal performance, the chunk size should be tuned to ensure sufficient parallelism while avoiding excessive memory usage (which can lead to contention) and minimizing the occurrence of a small final chunk (which can degrade performance due to underutilization).

{\footnotesize
\begin{verbatim}
pair_style mtp path/to/mtp/file
pair_style mtp/kk path/to/mtp/file chunksize 32768
pair_style mtp/small/kk path/to/mtp/file chunksize 32768
\end{verbatim}
}

The mode, either configuration or neighborhood, is read from the MTP file. For neighborhood active learning variations, we support both LAMMPS-like and MLIP-3-like processing of extrapolation grades. In the former style, the \code{pair\_style} is invoked as in inference.

{\footnotesize
\begin{verbatim}
pair_style mtp/extrapolation path/to/mtp/file
pair_style mtp/extrapolation/kk path/to/mtp/file chunksize 32768
pair_style mtp/extrapolation/small/kk path/to/mtp/file chunksize 32768
\end{verbatim}
}

A fix is then required to request extrapolation grades every \code{X} timesteps. 

{\footnotesize
\begin{verbatim}
fix mtp_grade all pair X mtp/extrapolation extrapolation 1
fix mtp_grade all pair X mtp/extrapolation/kk extrapolation 1
fix mtp_grade all pair X mtp/extrapolation/small/kk extrapolation 1
\end{verbatim}
}

The neighborhood extrapolation grades can then be accessed through the \code{f\_mtp\_grade} variable. LAMMPS's \code{dump} can then be used to periodically write the grades, and other desired per-atom properties to a file. Notably, if the user attempts to access grades on timesteps where extrapolation is not being calculated, the values will not be up-to-date.

{\footnotesize
\begin{verbatim}
dump my_dump all custom X path/to/dump f_mtp_grade 
\end{verbatim}
}

In the MLIP-3 style, the user specifies in order, the MTP file, the output file, the selection threshold, and the break threshold. Extrapolation is evaluated every timestep, and should the maximum grade surpass the selection threshold, the current configuration is written to the output file in the MLIP-3 format. Should this maximum grade surpass the break threshold, the simulation is immediately terminated. GPU variations still require the chunk size.

{\footnotesize
\begin{verbatim}
pair_style mtp/extrapolation path/to/mtp/file \
    path/to/output 2 10
pair_style mtp/extrapolation/kk path/to/mtp/file \ 
    path/to/output 2 10 chunksize 32768
pair_style mtp/extrapolation/small/kk path/to/mtp/file \
    path/to/output 2 10 chunksize 32768
\end{verbatim}
}

Configuration mode is only available with the MLIP-3 style. In either mode, the maximum extrapolation grade at each time step is available as a LAMMPS variable through a LAMMPS compute.

{\footnotesize
\begin{verbatim}
compute max_grade all pair mtp/extrapolation
compute max_grade all pair mtp/extrapolation/kk
compute max_grade all pair mtp/extrapolation/small/kk
\end{verbatim}
}

The variable can be accessed as usual through \code{c\_max\_grade[1]} and used in \code{fix halt}. Commonly, the user will print the grade along with other per-timestep quantities at regular intervals with LAMMPS \code{thermo}. Notably, if the user attempts to access this variable on timesteps where extrapolation is not being calculated, the value will not be up-to-date.

{\footnotesize
\begin{verbatim}
thermo_style custom step c_max_grade[1]
thermo X
\end{verbatim}
}

Much like some other MLIPs in LAMMPS, when invoking a LAMMPS script utilizing a MTP Kokkos GPU variation through the command line, additional flags are required:

{\footnotesize
\begin{verbatim}
-pk kokkos newton on neigh half
\end{verbatim}
}

A full example LAMMPS script and its command-line invocation are available in the supplementary materials \cite{zijian_2025_kokkos}.

\section{Illustrative examples}

We present three illustrative examples where GPU acceleration provides substantial benefits: (1) a large-scale simulation using a high-cost MTP, (2) a very large simulation with a medium-cost MTP, and (3) a medium-sized simulation that demonstrates active learning in practice. The LAMMPS input scripts for all examples are included in the supplementary materials \cite{zijian_2025_kokkos}. Visualizations were produced using OVITO~\cite{stukowski2009visualization}.

\subsection{Dislocation Glide in Silicon}
\begin{figure}[h!]
    \centering
    \adjustbox{valign=c}{\includegraphics[width=0.59\linewidth]{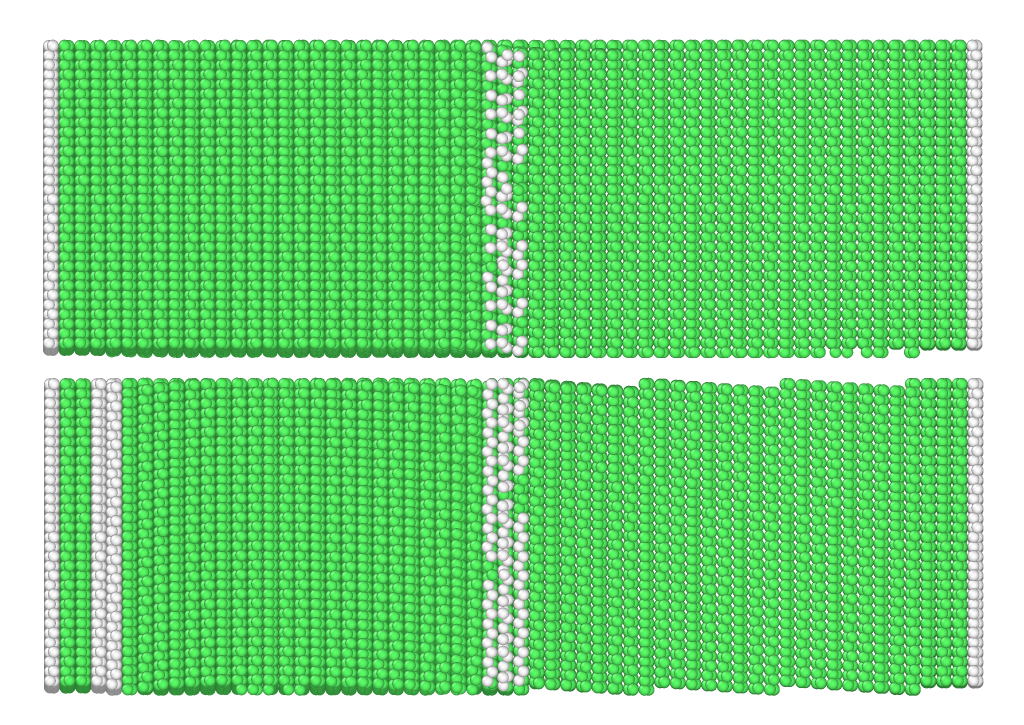}}%
    \adjustbox{valign=c}{\includegraphics[width=0.41\linewidth]{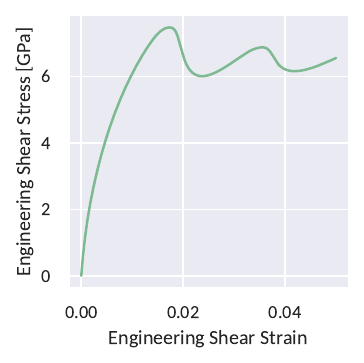}}
    \caption{Shearing of a C2 core type screw dislocation in silicon (115 thousand atoms, $5\times10^7$ s$^{-1}$ strain rate, 1 ns, 1 fs timestep). Top-Left: unstrained. Bottom-Left: strained. Right: engineering shear stress-strain curve.}
    \label{fig:shear}
\end{figure}

Dislocations are crystallographic line defects that disrupt the regular atomic structure of a material and often arise under mechanical stress or elevated temperatures. Once nucleated, they can propagate and multiply, affecting plasticity and deformation mechanisms. In semiconductors—critical to technologies such as transistors, LEDs, and solar cells—dislocations can influence fabrication processes, carrier transport, and overall device performance. Accurate simulation of dislocation behavior requires large simulation cells to capture long-range elastic fields and collective dislocation dynamics. As a case study, we generated a silicon cell containing approximately 115,000 atoms with a screw dislocation characterized by a C1-type core using Atomsk~\cite{hirel2015atomsk}. The structure was relaxed to the more stable C2 configuration, which is energetically favored in diamond cubic crystals. Molecular dynamics simulations were then performed using the isothermal–isobaric (NPT) ensemble and a level-26 silicon–oxygen MTP developed by Zongo et al.\cite{zongo2024unified}. A shear strain of $5\times10^7$ s$^{-1}$ was applied over a 1 ns simulation at 10 K to evaluate dislocation mobility. Results are shown in Figure \ref{fig:shear}.

Using the block-parallel inference variant on a full Narval GPU node, we achieved a simulation rate of 0.515 ns/day. In contrast, the original MLIP-3 implementation running on a full Narval CPU node achieved only 0.032 ns/day in a short trial run—a rate that is impractically slow for this type of simulation.

\subsection{Nanocrystalline Tension of Aluminum}
\begin{figure}[h!]
    \centering
    \adjustbox{valign=c}{\includegraphics[width=0.3\linewidth]{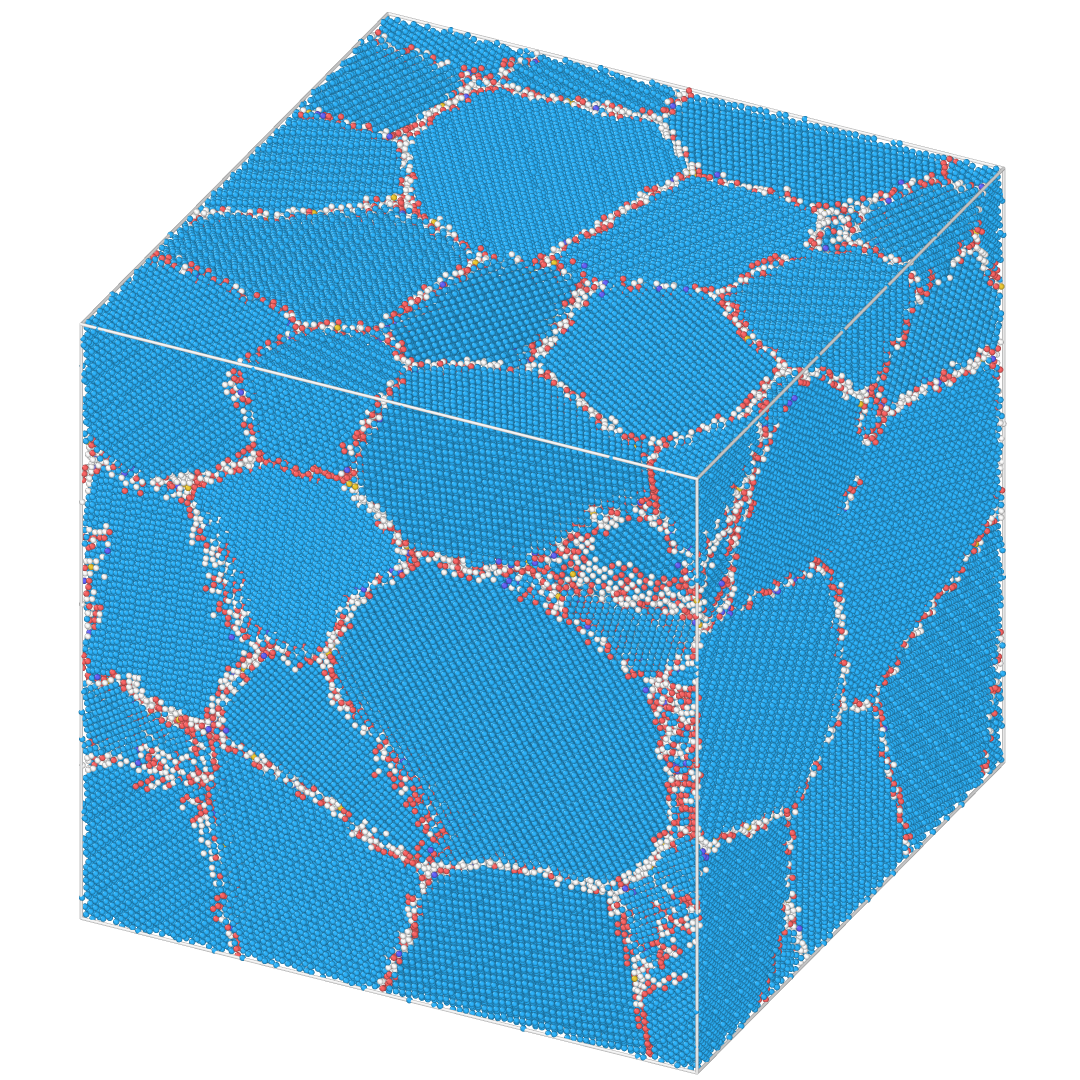}}%
    \adjustbox{valign=c}{\includegraphics[width=0.3\linewidth]{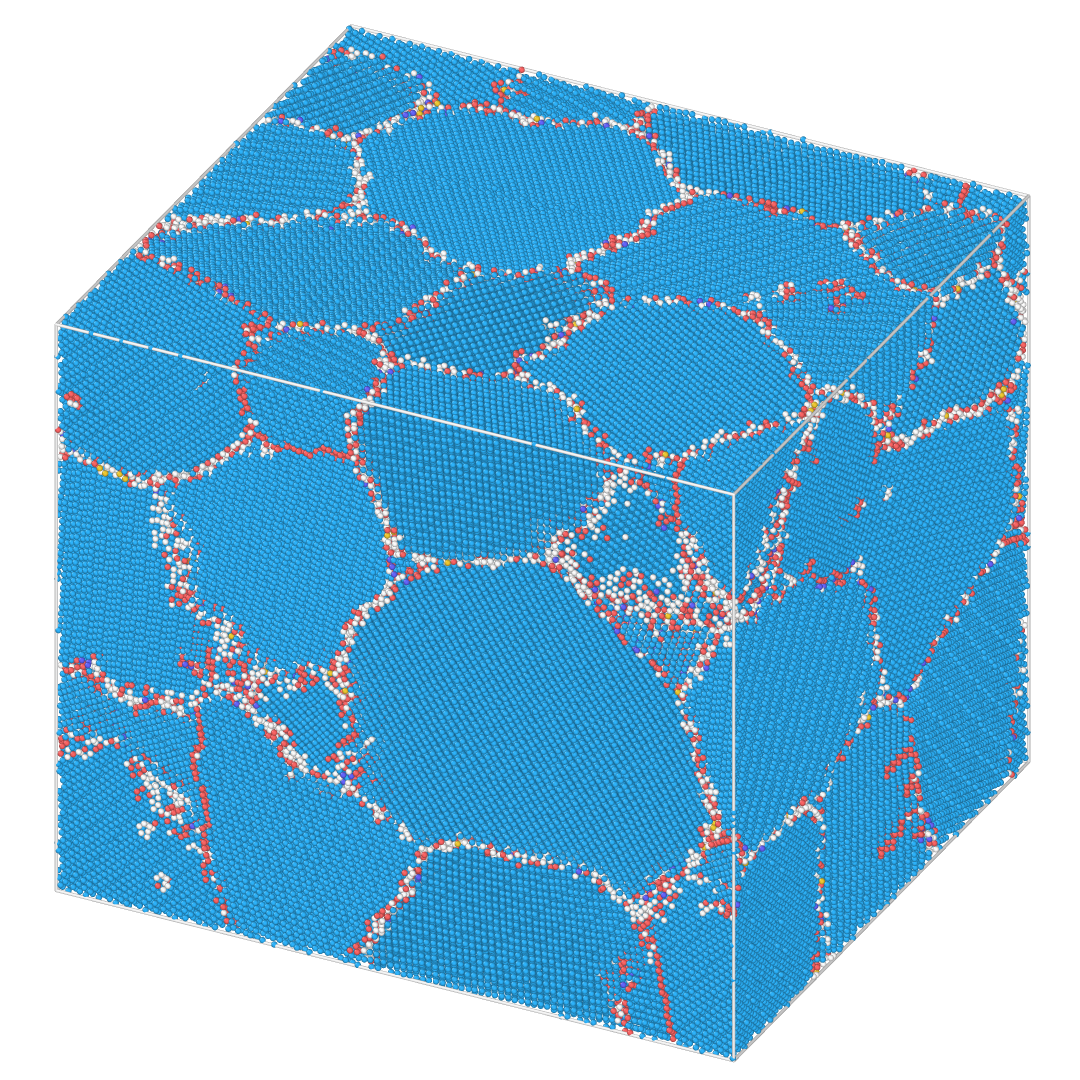}}%
    \adjustbox{valign=c}{\includegraphics[width=0.4\linewidth]{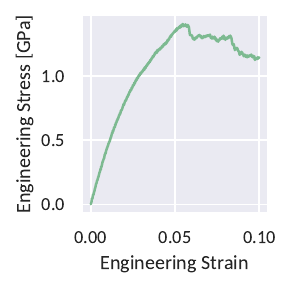}}
    \caption{Nanocrystalline tension of aluminum (1.00 million atoms, $10^8$ s$^{-1}$ strain rate, 1 ns, 1 fs timestep). Left: unstrained. Center: strained (0.1 strain). Right: engineering stress-strain curve.}
    \label{fig:nanocrystal}
\end{figure}

Nanocrystalline tension simulations are widely used to investigate defect dynamics, grain boundary behavior, and mechanical properties in materials with nanometer-scale grain sizes. To avoid artificial periodicity effects—where grains interact with their own periodic images—simulations must include a sufficiently large number of grains, often requiring millions of atoms for realistic structures. As a representative case, we generated an aluminum polycrystal using Atomsk~\cite{hirel2015atomsk} and employed a level-16 MTP from Novikov et al.\cite{novikov2020mlip}. The system contained one million atoms with a mean grain size of 11.1 nm. We performed uniaxial tensile deformation at 300K to a strain of 0.1 over 1 ns, corresponding to a strain rate of $10^8$ s$^{-1}$. Results are presented in Figure \ref{fig:nanocrystal}. Note that this potential was selected for demonstration purposes and was not specifically validated for this material or deformation mode.

Using the thread-parallel inference variant on a full Narval GPU node, we achieved a simulation rate of 0.453 ns/day. In comparison, the original MLIP-3 implementation on a full Narval CPU node yielded only 0.026 ns/day in a short trial run—a prohibitively slow rate for a simulation of this scale.

\subsection{Active Learning in a Coexistence Simulation}
\begin{figure}[h!]
	\centering 
	\includegraphics[width=0.7\linewidth]{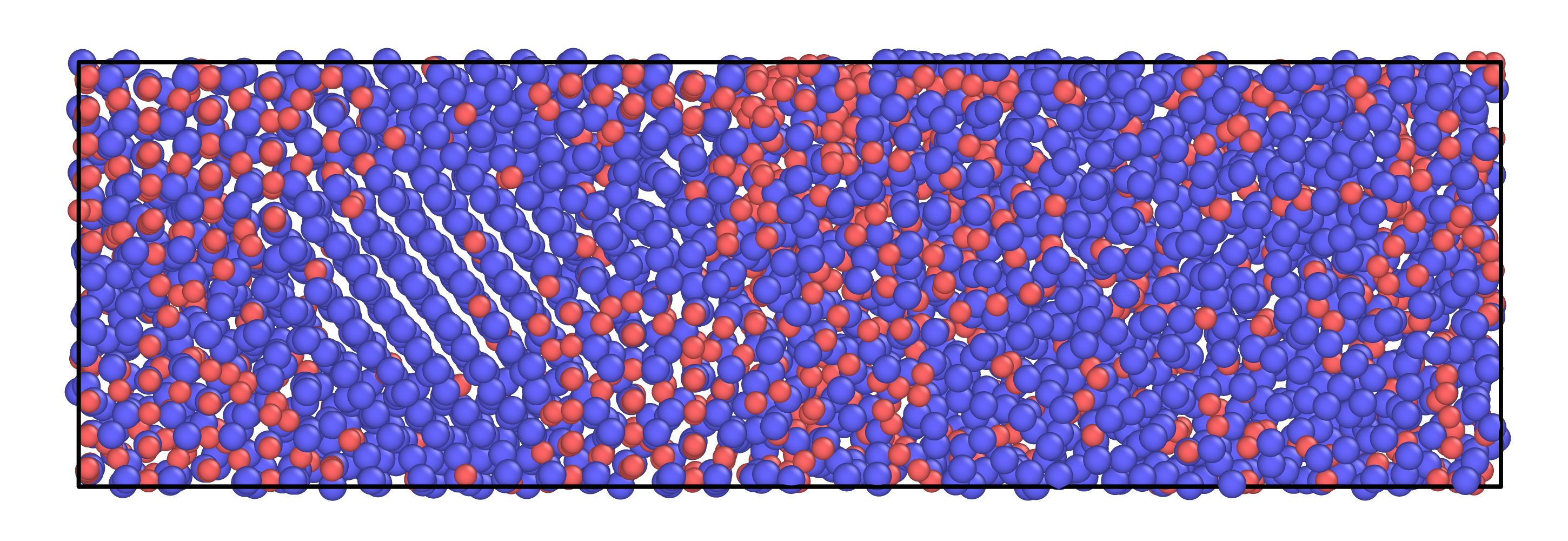}
	\caption{A eutectic sodium (red) and potassium (blue) solid-liquid interface which may be out-of-distribution and whose uncertainty was thus tested with active learning enabled. Solid (C14 + BCC) pictured left; liquid pictured right.} 
	\label{fig:interface}%
\end{figure}

After training an MTP on quantum-mechanical data, uncertainty quantification can be employed during early production MD simulations to assess the model’s reliability. In one such case, we developed a sodium–potassium alloy potential and sought to determine the eutectic melting point using a 3600-atom coexistence simulation featuring a solid–liquid interface (BCC + C14 + liquid), shown in Figure~\ref{fig:interface}. To ensure the level-18 MTP remained reliable when applied to potentially out-of-distribution configurations, we enabled active learning. This approach reduced the need to construct an excessively large validation set using quantum methods, while still assessing regions of potentially high extrapolation.

We performed this active learning simulation in configuration mode using both MLIP-3 on a full Narval CPU node and our block-parallel GPU implementation on a single A100. The MLIP-3 implementation achieved a simulation rate of 0.535 ns/day, while the GPU version reached 14.925 ns/day. Notably, despite the system size being well within the range typically amenable to MPI parallelization, MLIP-3's configuration mode exhibited limited scalability. This inefficiency would likely worsen if additional CPU nodes were used in an attempt to improve the simulation rate.


\section{Impact and Conclusions}
When exploring, discovering, and characterizing materials through atomistic simulations, many phenomena of interest require simulations involving millions of atoms. Examples include amorphous materials~\cite{madanchi2024future}, crack propagation~\cite{rountree2002atomistic}, nanocrystalline systems~\cite{zhou2014effects}, irradiation damage~\cite{nordlund2001defect}, and dislocation dynamics and plasticity~\cite{rao2017atomistic}. 

Despite its popularity, demonstrated successes, native support for active learning, and strong computational efficiency, the MTP has lacked GPU support—limiting its scalability on modern HPC systems. By introducing a family of GPU-accelerated variants, we significantly expand MTP's applicability to larger, more complex simulations. In particular, the block-parallel implementation achieves peak throughput with approximately 2000 atoms per GPU, enabling faster time-to-solution for many existing problems and making it feasible to deploy higher-level, more accurate MTPs that would otherwise be computationally prohibitive. These GPU implementations also enable medium-scale simulations on consumer-grade hardware or with Multi-Instance GPU (MIG), improving accessibility across a broader range of research environments. Additionally, our optimized CPU variant consistently outperforms the original implementation, with speedups of up to 2$\times$.

This software contribution is part of a collective effort to improve upon the MTP and other similar potentials such as the Equivariant Tensor Network Potential, the latter of which could be improved to support Kokkos.

\section*{Acknowledgements}
We would like to acknowledge Alexander Shapeev and Ivan Novikov for their insight and discussions regarding the MTP implementation and future work. 

\section*{Funding Sources}
This work was funded by the University Network for Excellence in Nuclear Engineering (UNENE), the Natural Sciences and Engineering Research Council of Canada (NSERC), and Mitacs. We thank the Digital Research Alliance of Canada (DRAC) for the generous allocation of computer resources. This work was partly funded by Atomic Energy of Canada Limited, Canada, under the auspices of the Federal Nuclear Science and Technology Program.

\section*{CRediT statement}
\textbf{Zijian Meng:} Conceptualization, Methodology, Software, Validation, Visualization, Investigation, Data Curation, Writing - Original Draft, Writing - Review \& Editing \textbf{Karim Zongo:} Validation, Writing - Review \& Editing \textbf{Edmanuel Torres:} Writing - Review \& Editing \textbf{Christopher Maxwell:} Writing - Review \& Editing \textbf{Ryan Grant:} Methodology, Resources, Writing - Review \& Editing, Supervision, Funding acquisition \textbf{Laurent Karim Béland:} Methodology, Resources, Writing - Review \& Editing, Supervision, Funding acquisition, Project administration

\section*{Declaration of generative AI and AI-assisted technologies in the writing process}
 During the preparation of this work, the authors used Gemini in order to help write plotting scripts in Python for benchmark data. After using this tool, the authors reviewed and edited the plots as needed and take full responsibility for the content of the published article.




\bibliographystyle{elsarticle-num} 

\bibliography{ref.bib}





\end{document}